\documentclass{iaus}
\usepackage{graphicx}

\title[JD 11.~~New observations of CP stars] 
{New observations of chemically peculiar stars with ESPaDOnS\footnote{Based on observations obtained at the Canada-France-Hawaii Telescope (CFHT)
       which is operated by the National Research Council of Canada, the Institut National des Sciences
       de l'Univers of the Centre National de la Recherche Scientique of France, and the University of Hawaii.}}

\author[Khalack et al.]   
{V. Khalack$^1$, B. Yameogo$^1$, C. Thibeault$^1$,
 F. LeBlanc$^1$}

\affiliation{$^1$Universit\'{e} de Moncton, Moncton, N.-B., Canada \\ email: {\tt khalakv@umoncton.ca}}

\pubyear{2013}
\volume{302}  
\pagerange{119--126}
\jname{Magnetic fields throughout stellar evolution}
\editors{A.C. Editor, B.D. Editor \& C.E. Editor, eds.}
\begin{document}

\maketitle

\begin{abstract}
We present the first results of the estimation of gravity and effective temperature for some poorly studied chemically peculiar stars
that were recently observed with the spectropolarimeter ESPaDOnS at CFHT. We have analyzed the spectra of HD71030, HD95608
and HD116235 to determine their radial velocity, $v\sin{i}$ and the average abundance of several chemical species.
We have also analyzed our results to verify for possible vertical abundance stratification of iron and chromium in these stars.

\keywords{Stars: chemically peculiar, stars: individual: (HD71030, HD95608, HD116235) }
\end{abstract}

\firstsection 
\section{Introduction}

Accumulation or depletion of chemical elements at certain optical depths brought about by atomic diffusion can modify the structure of stellar atmospheres and it is therefore important to gauge the intensity of such stratification. High resolution (R=65000) Stokes IV spectra of several CP stars with $v\sin{i} < 35$ km s$^{-1}$ were obtained recently with ESPaDOnS in the spectral domain from 3700\AA\, to 10000\AA.
Low rotational velocities of HD71030, HD95608 and HD116235 result in comparatively narrow and unblended line profiles, which are suitable for abundance analysis. They also help  produce a hydrodynamically stable atmosphere necessary for the diffusion process to take place.
To determine the effective temperature and gravity of these stars (see Table 1), the profiles of some Balmer and He\,{\sc i} lines were fitted with the help of FITSB2 code (\cite{Napiwotzki_etal04}) in the frame of synthetic fluxes calculated 
for different values of $T_{eff}$, $\log{g}$ and metallicity using the stellar atmosphere code PHOENIX (\cite{Hauschildt_etal97}). 


\begin{table}
  \begin{center}
  \caption{Parameters of stellar atmosphere and abundance for studied CP stars.}
  \label{tab1}
 {\scriptsize
  \begin{tabular}{|l|c|c|c|}\hline
{\bf Parameters } & {\bf HD71030 } & {\bf HD95608} & {\bf HD116235}  \\ \hline
$T_{eff}$,  K	  & 6780$\pm$200	&9200$\pm$200	&8900$\pm$200\\
$\log{g}$	  & 4.0$\pm$0.1	&4.2$\pm$0.2 & 4.3$\pm$0.1\\
$v\sin{i}$, km s$^{-1}$	& 9$\pm$2.0	& 17.2$\pm$2.0 & 20.2$\pm$2.0 \\
$v_{r}$, km s$^{-1}$& 38.1$\pm$1.0 	& -10.4$\pm$1.0 &-10.3$\pm$1.0 \\ \hline
$\log(FeI/N_{tot})$ &	-4.35$\pm$0.34(148)	&-3.95$\pm$0.35(139)	&-3.86$\pm$0.39 (126) \\
$\log(FeII/N_{tot})$&	-4.43$\pm$0.34 (23)	&-3.82$\pm$0.33 (32)	&-3.11$\pm$0.60 (18)\\
$\log(CrI/N_{tot})$ &	-6.11$\pm$0.45 (10)	&-4.66$\pm$1.20 (13)	&-5.06$\pm$0.89  (3)\\
$\log(CrII/N_{tot})$&	-6.27$\pm$0.09  (3)	&-5.26$\pm$0.59 (14)	&-5.14$\pm$0.61 (15)\\
$\log(NiI/N_{tot})$ &	-5.77$\pm$0.26 (40)	&-3.63$\pm$0.39 (11)	&-4.92$\pm$0.35 (10)\\
$\log(TiII/N_{tot})$&	-7.32$\pm$0.05  (2)	&-6.49$\pm$0.77 (35)	&-6.19$\pm$0.83  (6)\\ \hline
  \end{tabular}
  }
 \end{center}
\end{table}

\section{Spectral analysis and results}

Several spectra have been obtained for HD95608 and HD116235 during different nights. We have not detected any significant variability of line profiles and have combined the spectra to obtain one cumulative spectrum for each star taking into account the Doppler shift of the data due to the Earth's orbital motion.
In the case of HD71030, we have used a single spectrum, provided by ESPaDOnS, which has sufficient spectral resolution for abundance analysis.

The preliminary results of the abundance analysis are presented in Table 1 for HD71030, HD95608 and HD116235. The line profile simulation is performed using the ZEEMAN2 spectrum synthesis code (\cite{Landstreet88}) and LTE stellar atmosphere model calculated with PHOENIX (\cite{Hauschildt_etal97}) for the given values of effective temperature and gravity. For each element, we have selected a sample of unblended lines, which are clearly visible in the spectra. The element's abundance, radial velocity and $v\sin{i}$ were fitted using an automatic minimization routine (\cite{KhalackWade06}). For each ion presented in Table 1, the number of analyzed lines is specified in brackets.

\begin{figure*}[t]
\begin{center}
\includegraphics[width=1.8in,angle=-90]{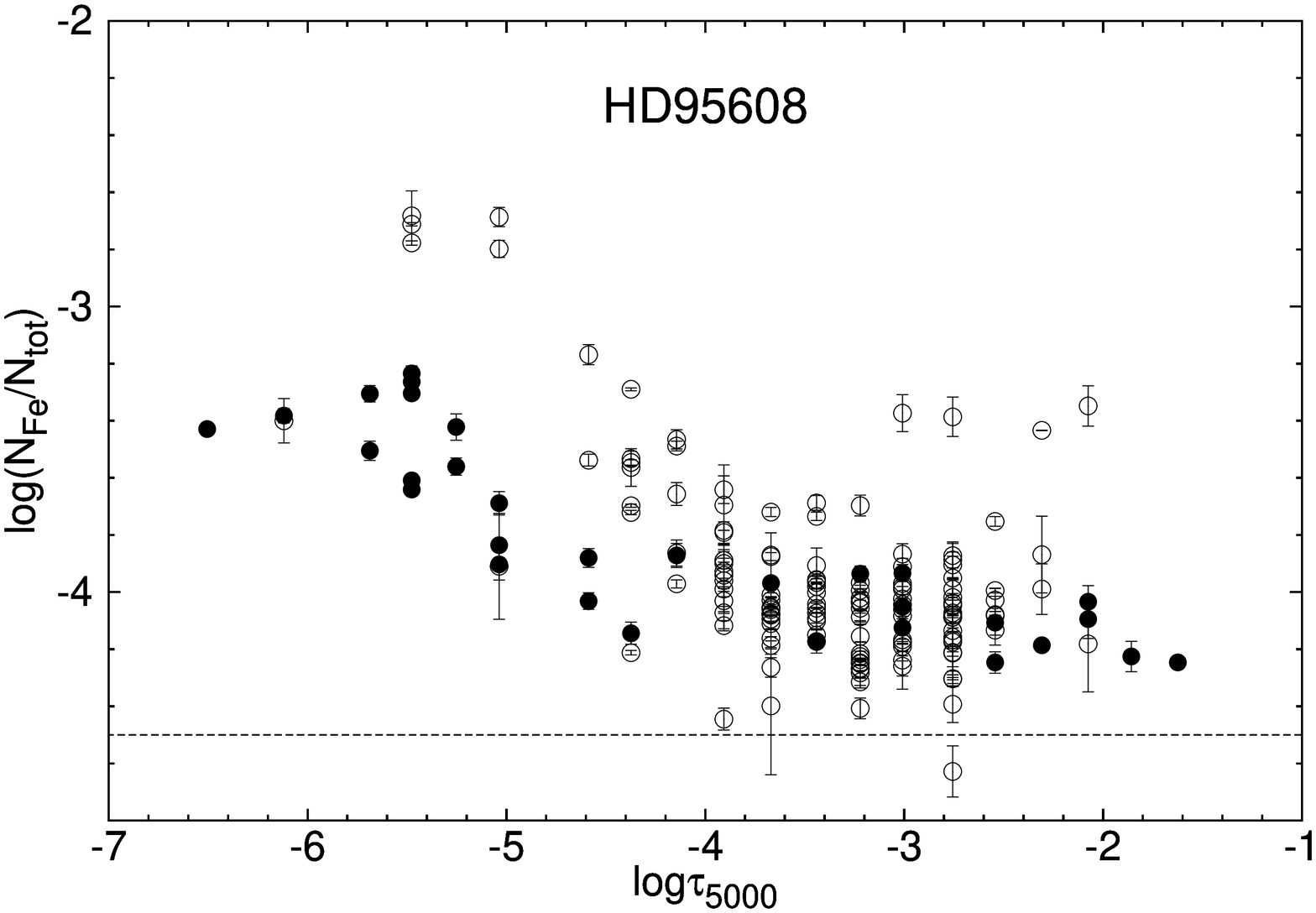}
\includegraphics[width=1.8in,angle=-90]{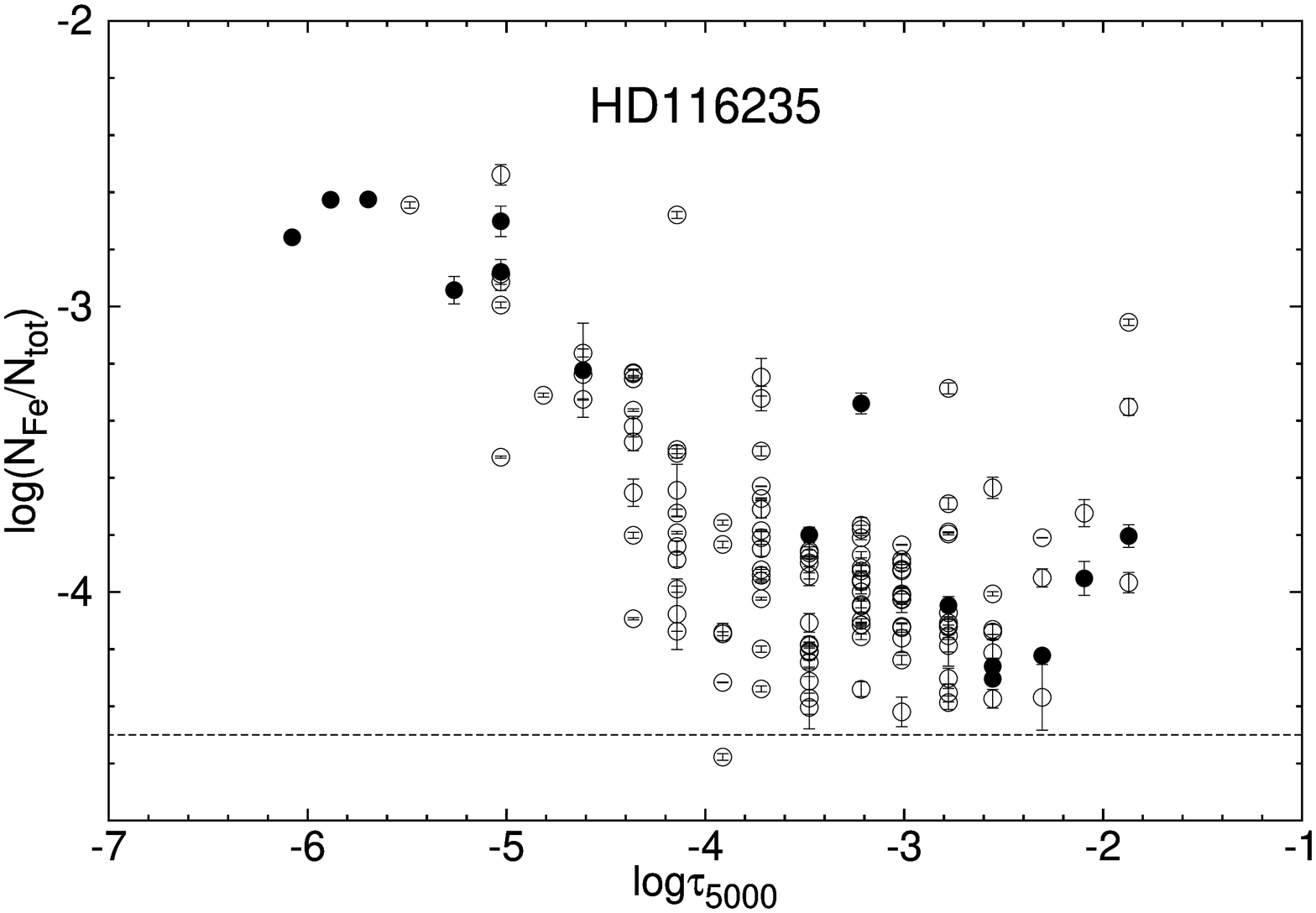}
 \caption{Variation of Fe abundance with optical depth in HD95608 and  HD116235. The results for Fe\,{\sc i} and Fe\,{\sc ii} lines are represented by open and filled circles, respectively. Each point provides the element's abundance determined from the analysis of a single line profile taking into account that its core is formed mainly at the given optical depth $\tau_{5000}$. The dashed line represents the solar abundance of iron.}
   \label{fig1}
\end{center}
\end{figure*}

For HD71030, HD95608 and HD116235, we have studied vertical stratification of iron and chromium in their atmospheres, taking into account a number of Fe\,{\sc i}, Fe\,{\sc ii}, Cr\,{\sc i} and Cr\,{\sc ii} lines visible in their spectra.  The results for Fe\,{\sc i} and Fe\,{\sc ii} lines are in good accordance amongst themselves for each of these three stars. Iron appears to be uniformly distributed in HD71030 and shows some signatures of abundance stratification with optical depth in HD95608 and HD116235 (see Fig.~\ref{fig1}). The chromium abundance seems to be constant relative to optical depth in HD71030, but the large scatter of derived chromium abundance does not allow us to determine if chromium is vertically stratified in HD95608 and HD116235.

\end{document}